\documentclass[11pt]{article}
\usepackage{moriond,epsfig}

\def\be{\begin{equation}}
\def\ee{\end{equation}}
\def\bea{\begin{eqnarray}}
\def\eea{\end{eqnarray}}

 \DeclareGraphicsExtensions{.pdf,.jpg,.jpeg}            
\usepackage{graphicx}                                  
 
\begin{document}
{\hfill \bf FERMILAB-CONF-12-144-E}\par
{\hfill \bf CDF/PUB/ELECTROWEAK/PUBLIC/10833}
\vspace*{4cm}
\title{Single $Z$ Production at the Tevatron}

\author{ Thomas J.~Phillips\\
for the CDF and D0 Collaborations }

\address{ Physics Department, Duke University, Durham, NC USA}

\maketitle\abstracts{
The production of single $Z$ bosons has been studied at  Fermilab's Tevatron by the CDF and D0 collaborations.  Measurements include the weak mixing angle, vector and axial-vector couplings between $Z$ bosons and light quarks, and angular coefficients in electronic decays which are sensitive to the spin of the gluon.  The collaborations have looked for and indication of new physics above the mass scale that can be directly produced at the Tevatron by studying the interference between $Z$ and photon propagators.  All measurements are consistent with Standard Model expectations.
}

\section{Introduction}

	The $Z$ boson provides a very clean system for studying electroweak physics and for searching for new physics because the electronic and muonic decays of the $Z$ can be identified with very little background.  The Tevatron now has substantial datasets of these decays, and these datasets have been used to measure a number of Standard-Model parameters and to look for new physics beyond the Standard Model (SM).  Here we describe a number of these measurements that have recently been completed.

\subsection{Forward-Backward Asymmetry}

\begin{figure}
  \begin{center}
    \includegraphics[width=0.49\textwidth]{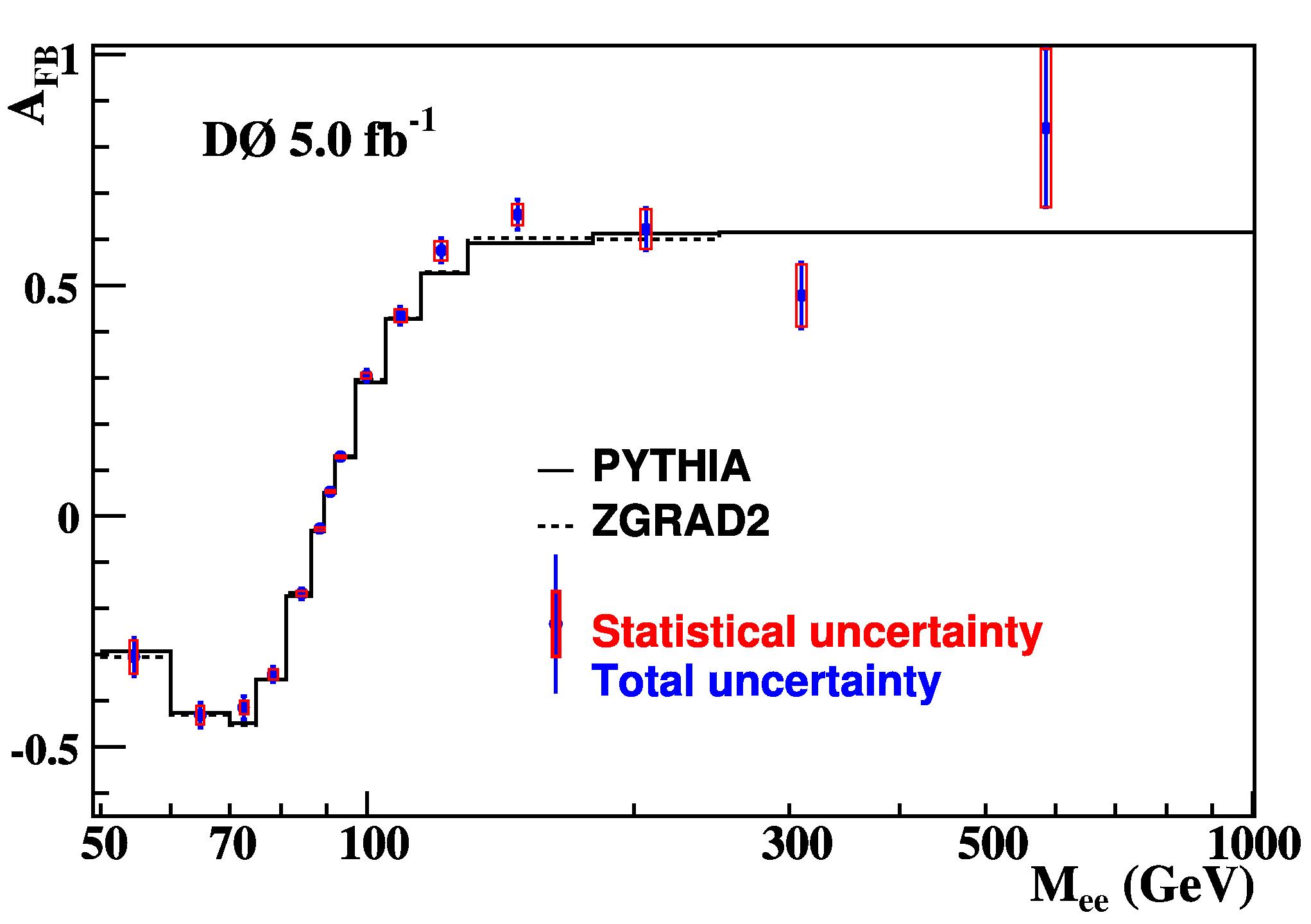}
     \includegraphics[width=0.49\textwidth]{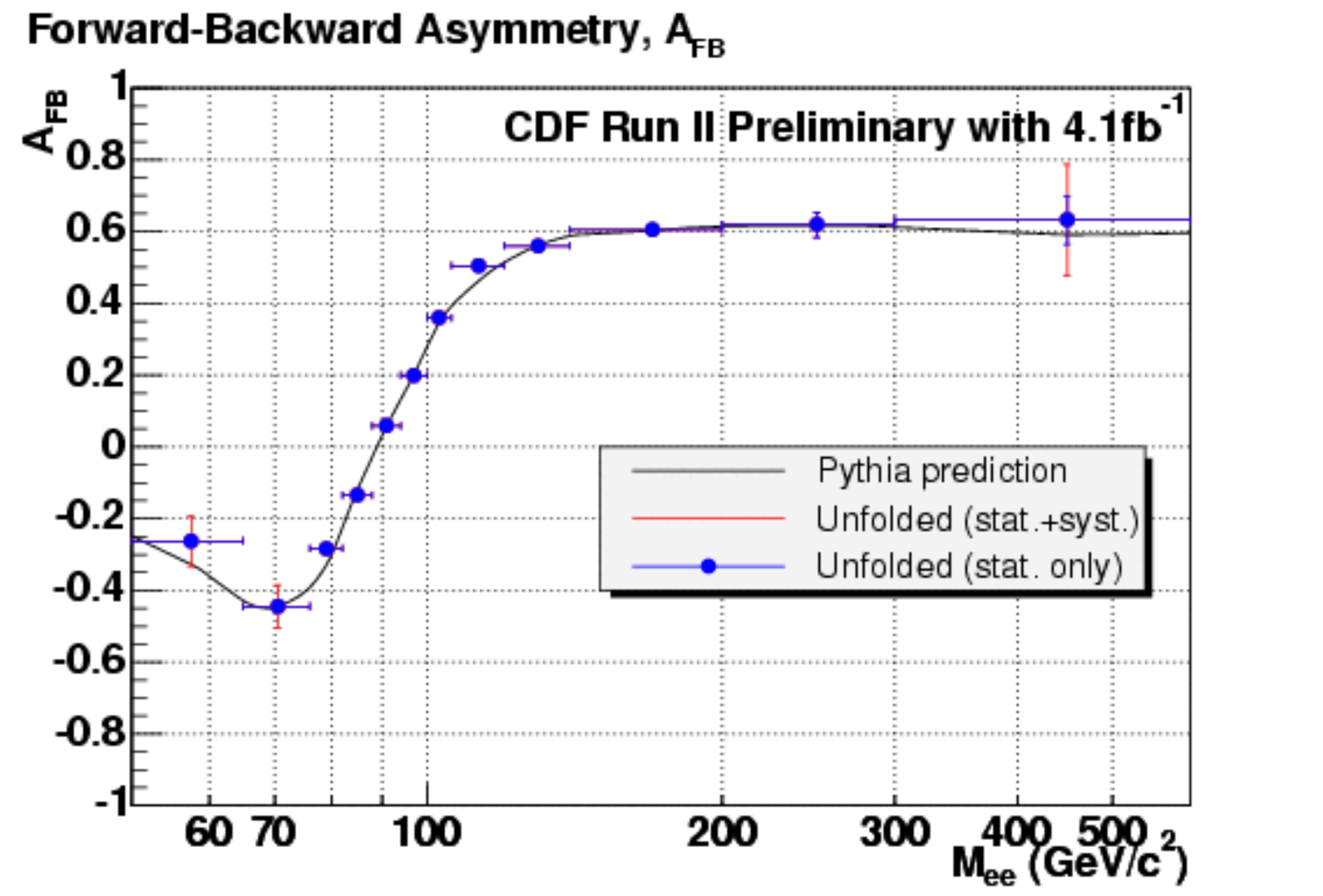}
     \end{center}
    \caption{Forward-Backward asymmetry as a function of dielectron invariant mass.  The left plot is the measurement from the D0 collaboration, and the right plot is the measurement from the CDF collaboration. \label{fig:Afb}}
\end{figure}

	Interference between $Z$-boson and photon propagators affect the direction of the electron resulting from the Drell-Yan process $p\bar{p} \to e^+e^-$.  One way to quantify this effect is to measure the forward-backward asymmetry of the electron.  This asymmetry is measured in the Collins-Soper frame~\cite{CollinsSoper},  which is defined as the rest frame of the electron pair, with the $z$ axis defined as the bisector of the directions of the incoming proton beam and the negative of the incoming antiproton beam.  Forward is defined as $\cos\theta^* > 0$ where $\theta^*$ is the angle between the negative electron's direction and the positive $z$ axis.  Backward is when $\cos\theta^* < 0$.  The measured asymmetry as a function of dielectron invariant mass is shown in Figure~\ref{fig:Afb} for both the D0 and the CDF collaborations.  
	
	The D0 collaboration has used this forward-backward asymmetry $A_{FB}$ measurement to extract the value of the weak mixing angle $\sin^2\theta^l_{eff} = 0.2309 \pm 0.0008 \pm 0.0006$~\cite{D0_ZAsym}.  This measurement is dominated by the high statistics in the $Z$ pole region, and it is consistent with other measurements of this quantity which average to $0.23153\pm 0.00016$.  
	
\begin{figure}
  \begin{center}
    \includegraphics[width=0.49\textwidth]{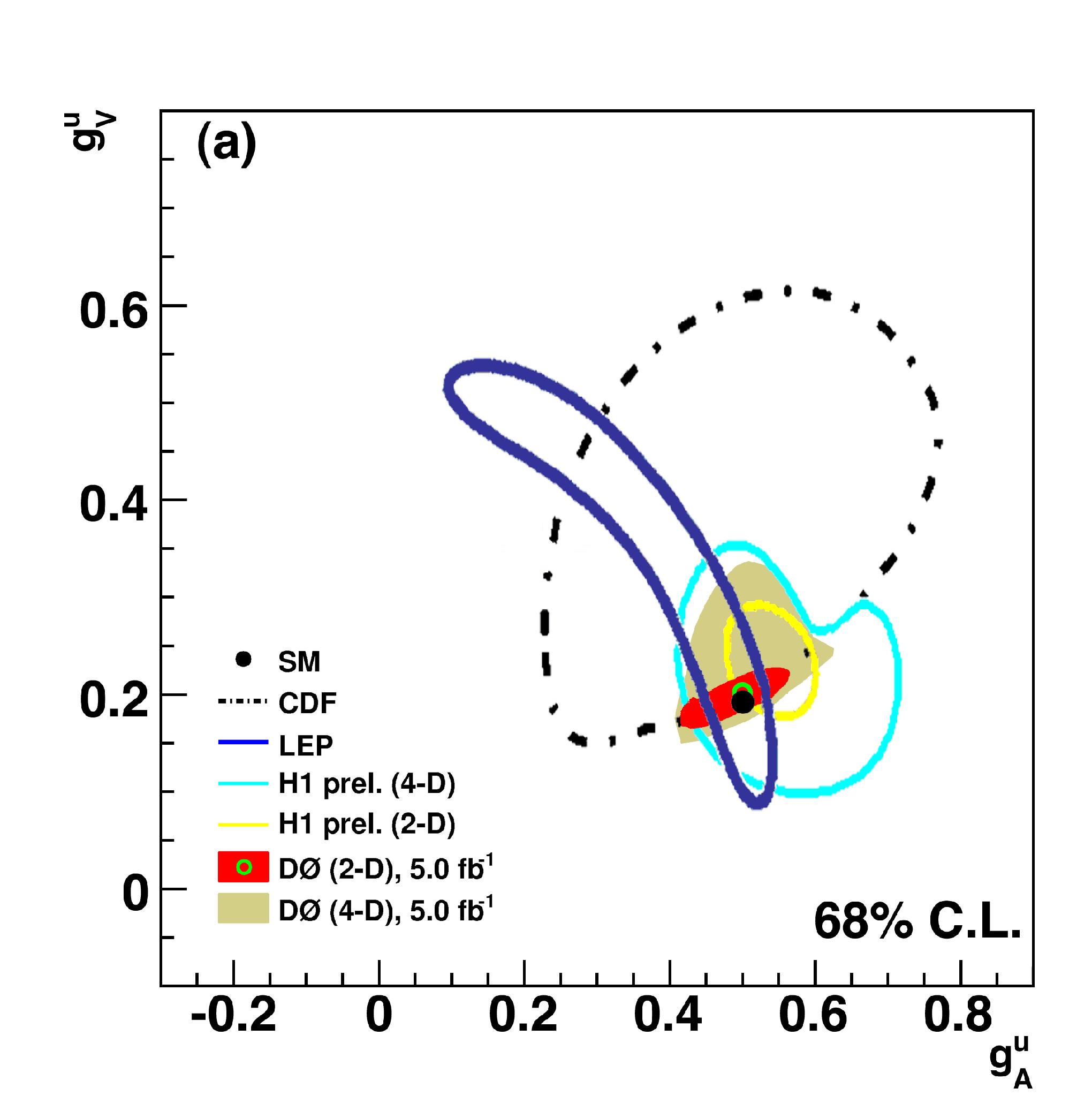}
    \includegraphics[width=0.49\textwidth]{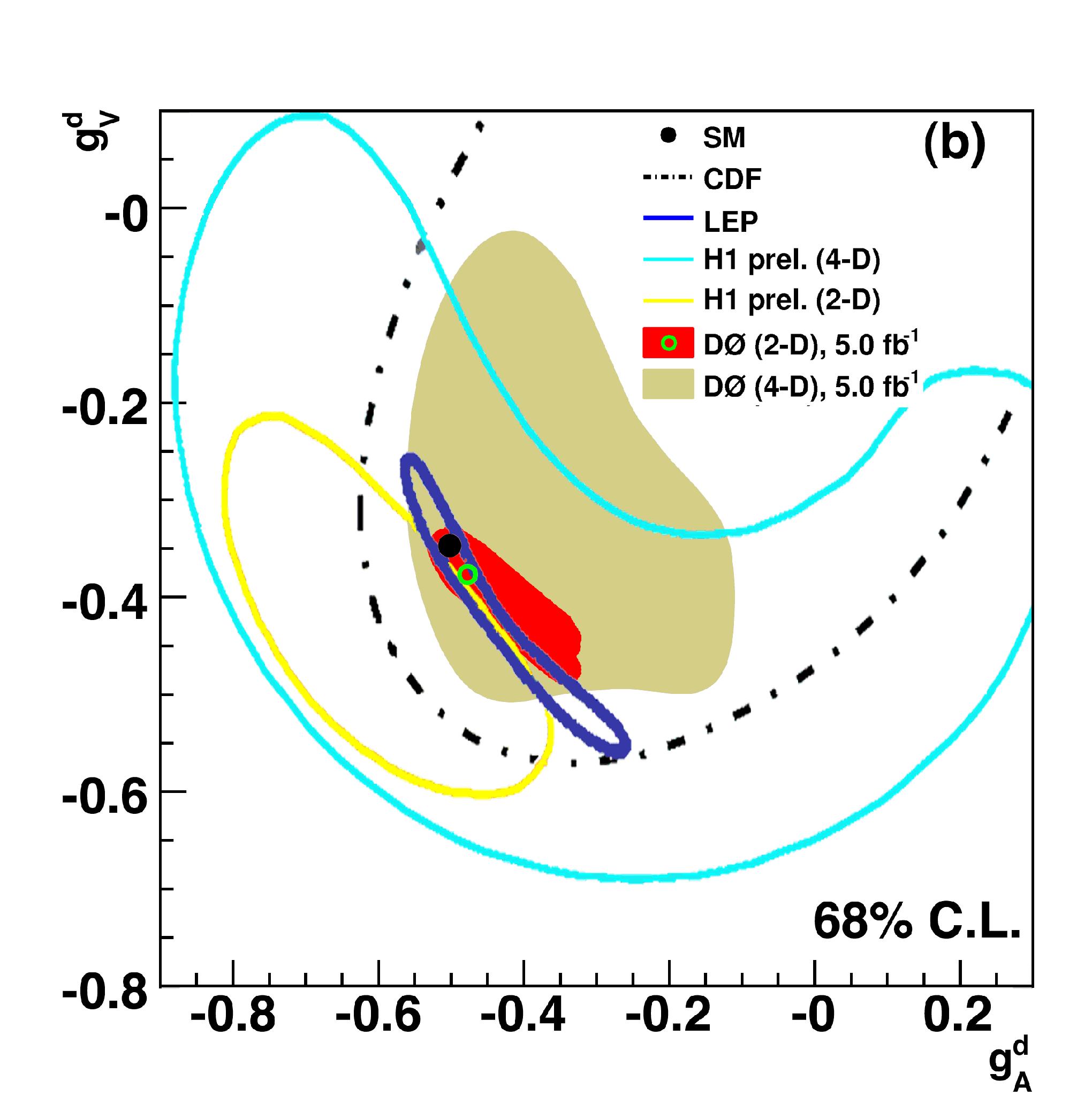}
  \end{center}
    \caption{Sixty-eight percent C.L. limits on couplings between $Z$ bosons and up quarks (a) and down quarks (b). Limits are shown for both 2-D and 4-D fits, as well as the best fit 2-D values and the SM expected values, \label{fig:D0CouplingLimits}}
\end{figure}

		The direction of the electron is affected by whether the $Z/\gamma$ propagator was produced by an up or a down quark, so the D0 collaboration has used their asymmetry data to extract limits on the vector ($g_V$) and axial-vector ($g_A$) couplings of light quarks to the $Z$ boson~\cite{D0_ZAsym}.  They used RESBOS~\cite{RESBOS} to make templates using different values of the couplings while the value of the coupling between electrons and $Z$ bosons was fixed to its SM value and $\sin^2\theta^l_{eff}$ was set to its global average.  The results are shown in Figure~\ref{fig:D0CouplingLimits}, which shows the 68\% C.L. allowed regions for both 2-dimensional fits (where the couplings for the other light quark are fixed at their SM values) and 4-dimensional fits, where all the light-quark couplings are allowed to vary at the same time.  All the fits are consistent with SM expectations.

	$A_{FB}$ is sensitive to new physics at masses higher than could be directly produced at the Tevatron.  In the absence of new physics, $A_{FB}$ is expected to be approximately constant at a value of 0.6 for dielectron invariant masses substantially above the $Z$ mass.  New physics could interfere with the $Z\gamma$ propagator and change this value.  Both the D0 and CDF $A_{FB}$ measurements (shown in Figure~\ref{fig:Afb}) are consistent with SM expectations, and therefore limit the possibility of new physics such as a massive $Z^\prime$ that interferes with the SM propagators.

\section{Angular Distributions}

The CDF collaboration has chosen to measure $\sin^2\theta_W$ form the angular distributions of the final-state electron rather than from the forward-backward asymmetry.  The general expression for the angular distribution of the final-state electron in the Collins-Soper frame~\cite{CollinsSoper} is:
\begin{eqnarray*}
{d\sigma\over {d\cos\theta \, d\phi}} & \propto & (1+\cos^2\theta) \\
& + & {1 \over 2} A_0(1 - 3\cos^2\theta) + A_1\sin 2\theta\cos\phi\\
& + & {1\over 2} A_2\sin^2\theta\cos 2\phi + A_3\sin\theta\cos\phi\\
& + & A_4\cos\theta + A_5\sin^2\theta\sin2\phi\\
& + & A_6\sin 2\theta\sin\phi + A_7 \sin\theta\sin\phi \\
\end{eqnarray*}
Here the coefficients $A_0$ to $A_7$ are functions of the dielectron mass $M_{ee}$, the transverse momentum $P_T$ of the $Z$ boson, and the rapidity $y$.  In perturbative QCD, $A_5$, $A_6$, and $A_7$ are near 0, $A_1$ and $A_3$ are small when integrated over $\pm y$, $A_4$ is sensitive to $\sin^2\theta_W$, and $A_0 = A_2$.  This last expression is the Lam-Tung equation,  and it is only valid for spin-1 gluons.  Since this expression is badly broken for spin-0 gluons, verifying that $A_0 = A_2$ provides evidence for spin-1 gluons.

\begin{figure}
  \begin{center}
    \includegraphics[width=0.49\textwidth]{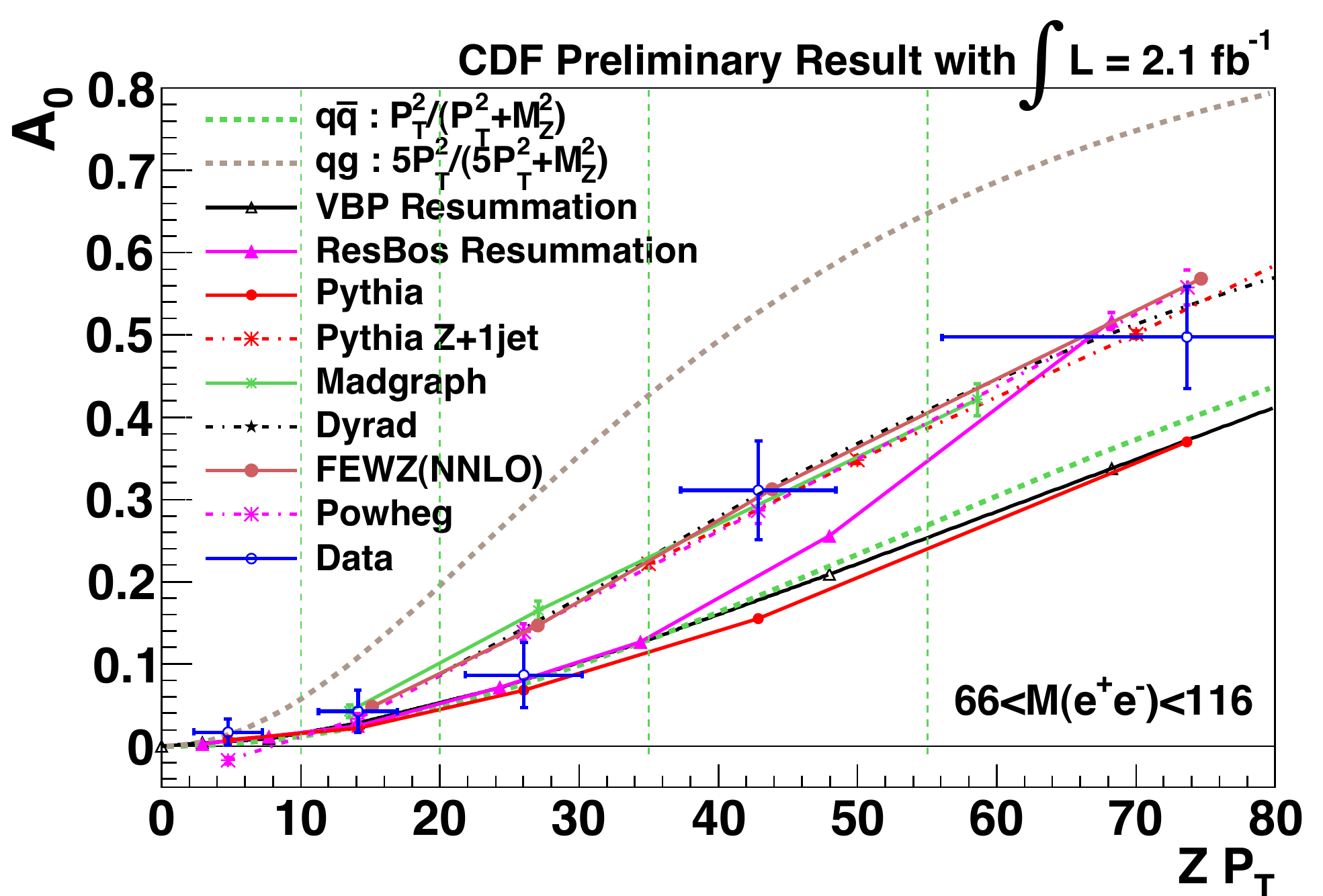}
    \includegraphics[width=0.49\textwidth]{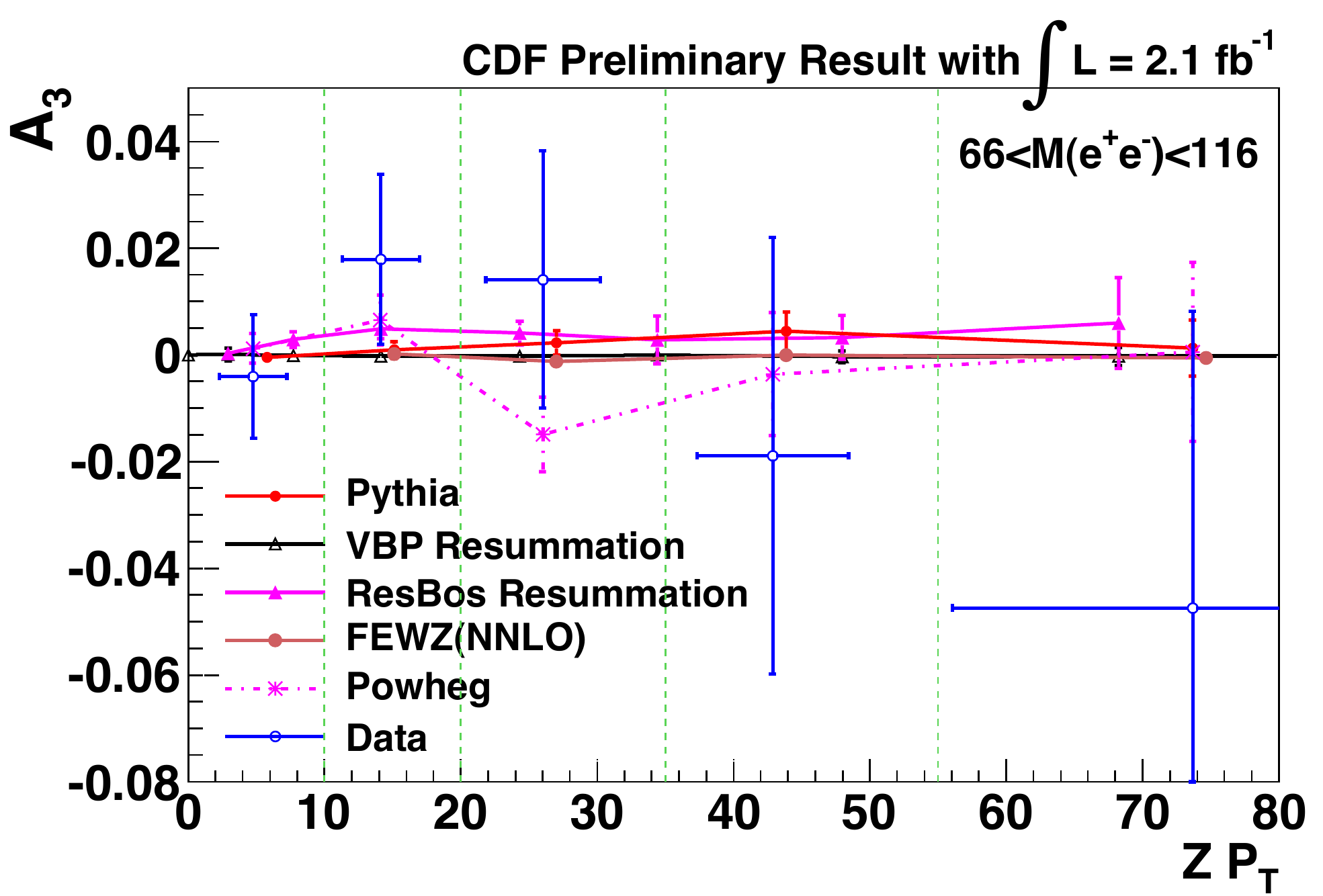}
    \includegraphics[width=0.49\textwidth]{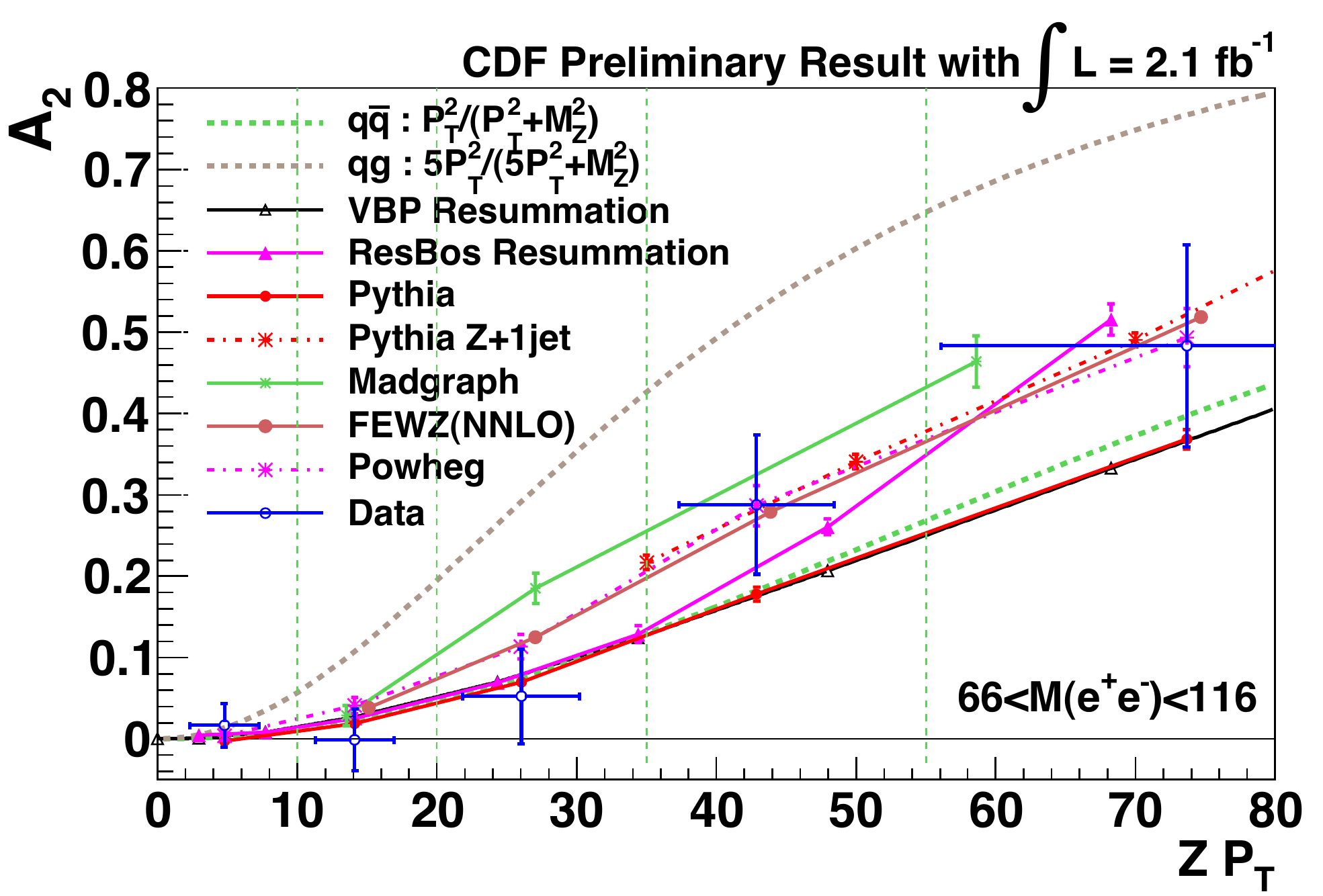}
    \includegraphics[width=0.49\textwidth]{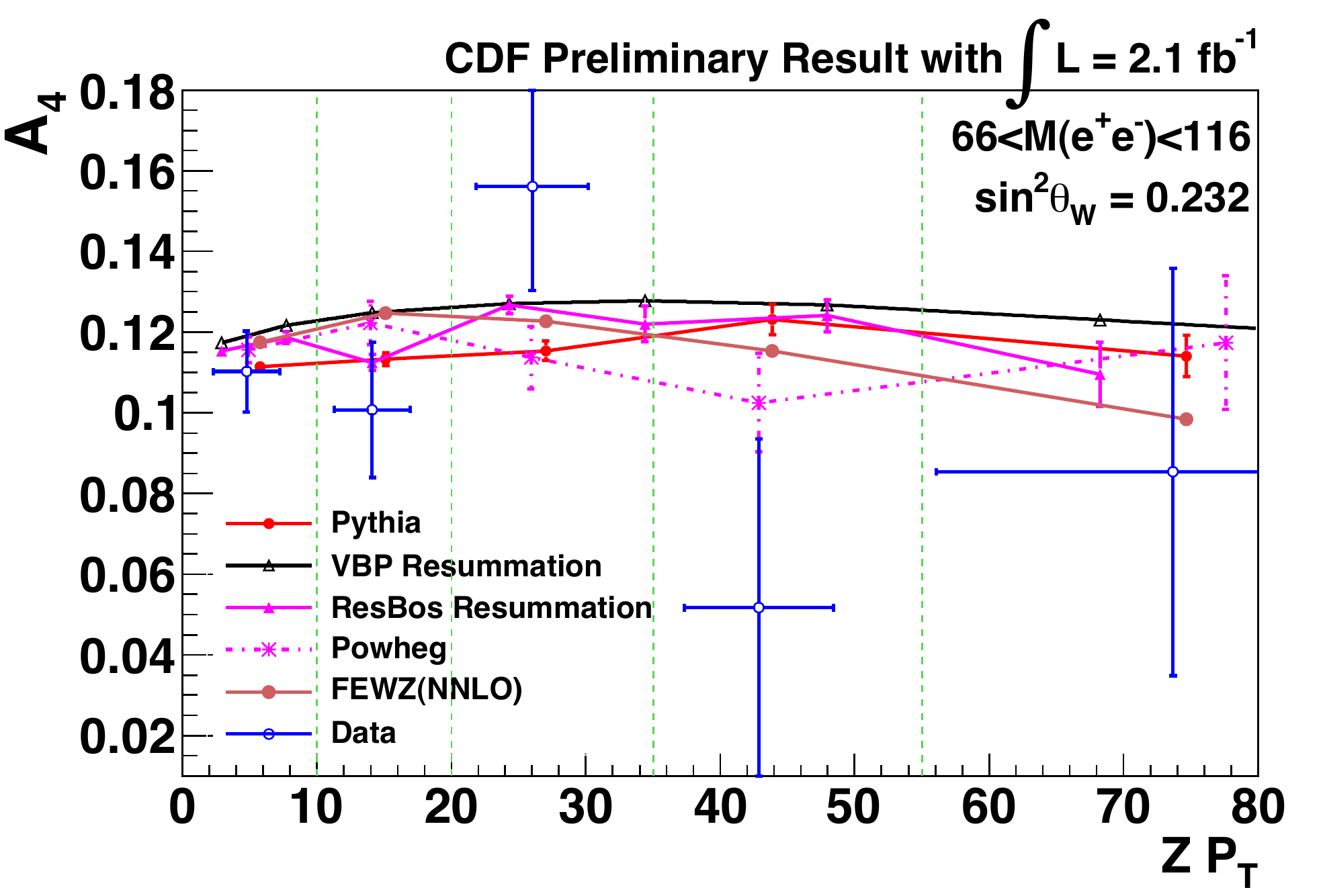}
  \end{center}
        \caption{Coefficients for the angular distribution of the final-state electron as measured by the CDF collaboration.  The data are the crosses, and the various lines correspond to different theoretical expectations. \label{fig:CDF_Zangular}}
\end{figure}

	The measured angular coefficients~\cite{CDF_ZAngular} are plotted as a function of $P_T$ in Figure~\ref{fig:CDF_Zangular}.  We see that these measurements are consistent with the Lam-Tung expression and therefore with a spin-1 gluon.  $A_3$ is close to 0 as expected for pQCD, and $A_4$ is used to extract $\sin^2\theta_W = 0.2329 \pm 0.0008 ^{+0.0010}_{-0.0009}$(QCD).  All these results are consistent with SM expectations.

\section{$Z$ Transverse Momentum}

\begin{figure}
  \begin{center}
    \includegraphics[width=0.9\textwidth]{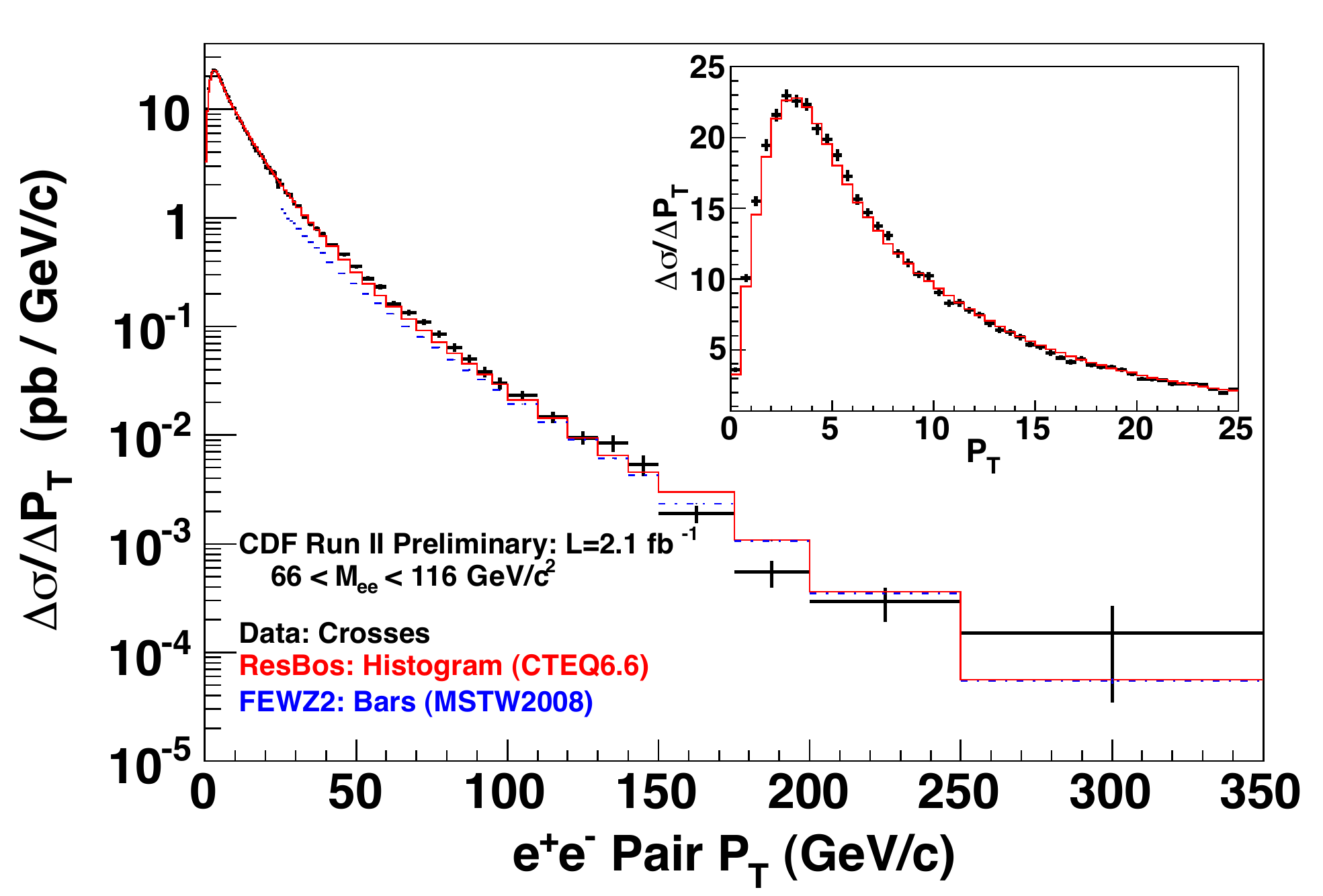}
  \end{center}
    \caption{Comparison of CDF data (crosses) to ResBos (solid) and NNLO (FEWZ2, dashed) theoretical calculations. The  data uncertainties include the statistical uncertainty for both the data and the unforlding,  and the 1\% efficiency measurement uncertainty, all combined in quadrature. Luminosity uncertainties are not included; the ResBos total cross section is 254 pb.
 \label{fig:zdsdptPRE}}
\end{figure}

	The same dataset used by CDF to measure the angular coefficients was used to measure the transverse momentum of the $Z$ boson.  At low $P_T$, the measurement smearing is large (on order of 2.2 GeV$/c$) compared to the bin size of 0.5 GeV$/c$.  The unfolding is done by first correcting the input $P_T$ distribution for the Pythia Monte Carlo generator~\cite{Pythia} until the ratio of data/simulation is flat, and then using the simulation to determine bin-by-bin unfolding.  
	
\begin{figure}
  \begin{center}
    \includegraphics[width=0.9\textwidth]{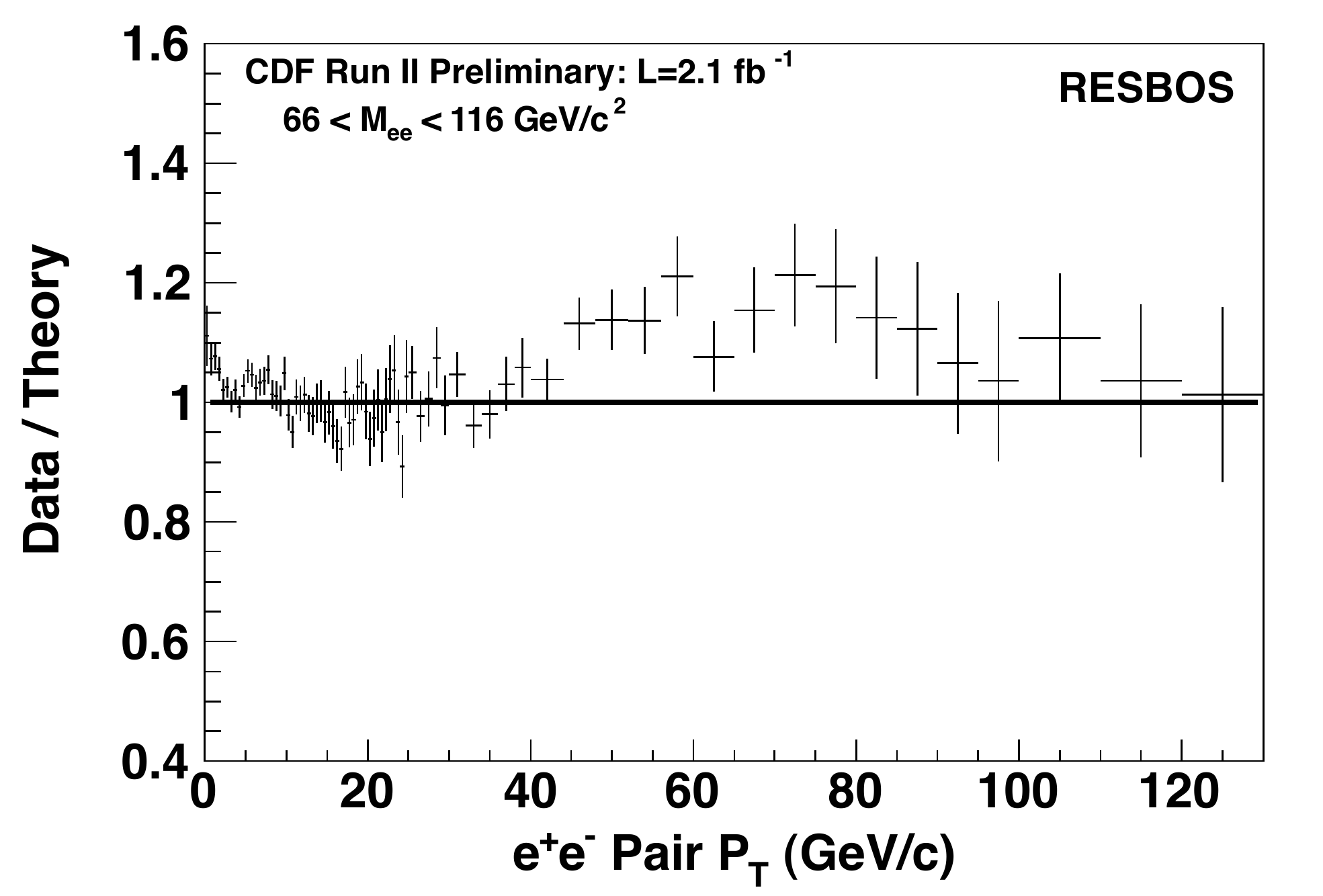}
  \end{center}
    \caption{Ratio of data to ResBos theory.
 \label{fig:dataOvrResb}}
\end{figure}

		The unfolded $P_T$ distribution is shown in Figure~\ref{fig:zdsdptPRE} along with the NNLO theoretical prediction (FEWZ2) and the resummation prediction (ResBos\cite{RESBOS}).  The NNLO prediction is consistent with the measured result for high $P_T$, while the ResBos prediction does a good job of matching the data over the entire $P_T$ range.  A closer look at the ResBos prediction is shown in Figure~\ref{fig:dataOvrResb}, which plots the ratio of the data to the ResBos theory.  THe deviation seen in the region $40 < P_T < 90$ GeV$/c$ is where ResBos resummed, asymptotic, and perturbative cross sections are matched.  Apparently the modeling could be improved in this region.

\section{Conclusions}

	The Tevatron experiments CDF and D0 have substantial datasets of well-identified $Z$ bosons.  These datasets have been used to measure Standard Model parameters including the weak mixing angle (D0: $0.2309\pm 0.0008 \pm 0.0006$; CDF: $0.2329 \pm 0.0008^{+0.0010}_{-0.0009}$(QCD)).  Measurements are consistent with Standard Model expectations, such as spin-1 gluons rather than spin-0 gluons.  Measurements of the $Z$-boson $P_T$ are now precise enough to help refine Drell-Yan phenomenology.  

\section*{Acknowledgments}
We thank the Fermilab staff and the technical staffs of the participating institutions for their vital contributions. This work was supported by the U.S. Department of Energy and National Science Foundation; the Italian Istituto Nazionale di Fisica Nucleare; the Ministry of Education, Culture, Sports, Science and Technology of Japan; the Natural Sciences and Engineering Research Council of Canada; the National Science Council of the Republic of China; the Swiss National Science Foundation; the A.P. Sloan Foundation; the Bundesministerium f\"ur Bildung und Forschung, Germany; the Korean World Class University Program, the National Research Foundation of Korea; the Science and Technology Facilities Council and the Royal Society, UK; the Russian Foundation for Basic Research; the Ministerio de Ciencia e Innovaci\'{o}n, and Programa Consolider-Ingenio 2010, Spain; the Slovak R\&D Agency; the Academy of Finland; and the Australian Research Council (ARC).

\end{document}